\documentclass[journal]{IEEEtran}
\pdfoutput=1

\ifCLASSINFOpdf
\else
\fi
\usepackage{amsmath}
\usepackage{nth}

\usepackage{algorithmic}

 \usepackage[caption=false,font=footnotesize]{subfig}
\usepackage{url}

\usepackage{standalone}

\usepackage{hyperref}

\usepackage{soul}

\usepackage{textcomp}

\usepackage{todonotes}

\usepackage{bm}
\newcommand{\mat}[1]{\bm{#1}}

\usepackage{commath}
\usepackage{amsopn}

\DeclareMathOperator{\E}{E}

\usepackage{stmaryrd}

\usepackage{mathtools}

\DeclarePairedDelimiter{\ct}{\llbracket}{\rrbracket}
\DeclarePairedDelimiter{\pdct}{[}{]}

\usepackage{tikz}

\usetikzlibrary{positioning,calc,shapes,arrows,shapes.geometric,intersections}

\usepackage{newfloat}
\DeclareFloatingEnvironment{crypto}

\usepackage[
  n,
  advantage,
  operators,
  sets,
  adversary,
  landau,
  probability,
  notions,
  logic,
  ff,
  mm,
  primitives,
  events,
  complexity,
  asymptotics,
  keys]{cryptocode}

\usepackage{acronym}
\acrodef{FRR}{False Reject Rate}
\acrodef{FAR}{False Accept Rate}
\acrodef{GAR}{Genuine Accept Rate}
\acrodef{EER}{Equal Error Rate}
\acrodef{ROC}{Receiver Operation Characteristic}

\acrodef{LR}{Likelihood Ratio}
\acrodef{LLR}{Log-Likelihood Ratio}
\acrodef{SVM}{Support Vector Machine}
\acrodef{GC}{Garbled Circuit}

\acrodef{PDF}{probability density function}

\acrodef{DLP}{discrete log problem}
\acrodef{DDH}{Decisional Diffie-Hellman}
\acrodef{INDCPA}[$\indcpa$]{Indistinguishable Under Chosen Plaintext Attack}

\begin{document}
%
\title{Fast and Accurate Likelihood Ratio Based Biometric Comparison in the Encrypted Domain}
%
%
%


\author{Joep~Peeters,~\IEEEmembership{University of Twente}
             Andreas~Peter,~\IEEEmembership{University of Twente}\\
             Raymond N. J. Veldhuis,~\IEEEmembership{University of Twente}}
\maketitle

\begin{abstract}
As applications of biometric verification proliferate, users become more vulnerable to privacy infringement.
Biometric data is very privacy sensitive as it may contain information as gender, ethnicity and health conditions which should not be shared with third parties during the verification process.
Moreover, biometric data that has fallen into the wrong hands often leads to identity theft.
\textit{\textbf{Secure}} biometric verification schemes try to overcome such privacy threats. Unfortunately, existing secure solutions either introduce a heavy computational or communication overhead or have to accept a high loss in accuracy; both of which make them impractical in real-world settings.
This paper presents a novel approach to secure biometric verification aiming at a practical trade-off between efficiency and accuracy, while guaranteeing full security against honest-but-curious adversaries.
The system performs verification in the encrypted domain using elliptic curve based homomorphic ElGamal encryption for high efficiency.
Classification is based on a log-likelihood ratio classifier which has proven to be very accurate.
No private information is leaked during the verification process using a two-party secure protocol.
Initial tests show highly accurate results that have been computed within milliseconds range.
\end{abstract}


%
\IEEEpeerreviewmaketitle

\section{Introduction}\label{sec:introduction}
Biometrics are discriminative characteristics of the human body that can be used for authentication (i.e.\ identity verification) or identification related questions.
These characteristics can be physiological aspects of the body (e.g.\ fingerprints) or can be embedded in the behavior of a person (e.g.\ a signature).
A biometric system uses these \emph{biometric identifiers} to classify an individual to an earlier enrolled identity in the system.

Biometric authentication relies on the primitive ``user \emph{is} something''.
Alternatives are ``user \emph{knows} something'' (e.g.\ passwords) or ``user \emph{has} something'' (e.g.\ key cards).
Each primitive has its own upsides and its downsides.
The main merit of biometrics is their convenience since there is no need to remember anything and it simply is available when needed.
Though, the main drawback is that biometric identifiers are usually hard to keep secret and are not replaceable~\cite{Schneier1999}.
Also the fact that a biometric identifier contains privacy sensitive information about a person's gender, ethnicity and health implies that this data should be handled with care~\cite{Mordini, Ashbourn2014}.
As applications of biometric verification proliferate, users are requested to hand over their identity more frequently to third parties which might not be trustworthy.

In order to protect against this privacy infringement, biometric systems should be able to perform the classification without revealing the original biometric data.
Straightforward hashing techniques, as applied to passwords, are not applicable here as the classifier should be able to handle intra-user variations in the biometric captures.
An overview of related work in \autoref{sec:template-protection} shows that there are different methods to protect biometric data.
Some of these rely on one-way transformation functions to obfuscate the data, where others try to handle the intra-user variations using error correcting techniques.
The main problem with these approaches are the negative effects on the biometric classification accuracy.
Modern approaches use homomorphic encryption to perform the classification on encrypted versions of the biometric templates, but tend to be slow for highly accurate systems.

In this work we present a novel approach to perform a biometric classification on a centralized system designed with privacy protection in mind.
On top of that we aim to create a system with high accuracy \textit{and} high performance.
The system is build around a log-likelihood ratio classifier which is known to yield optimal results in terms of accuracy~\cite{Bazen2004}.
Privacy is protected by performing the classification in the encrypted domain using homomorphic encryption.
This allows us to protect both the inputs and the outputs of the verification process such that no privacy sensitive data is revealed.
We use an implementation of ElGamal on elliptic curves which allows us to perform fast classifications (millisecond range).
In order to perform the verification efficiently we quantize the classifier
without giving up too much of accuracy (an EER of 0,3\% versus 0,2\% in the optimal case).

As this paper focusses on privacy protection, the terms security and privacy might become unclear.
System security prevents an attacker from circumventing the system by getting accepted even though he cannot verify his identity.
Privacy covers the fact that a compromised system does not leak any information about the biometric data it stores.
These two principles overlap where an attacker tries to get accepted by means of a stolen biometric sample.
For instance by eavesdropping on an unsafe authentication attempt from a legit user.
Therefore it is important to protect the biometric data in order to strengthen the overall system security.

The remainder of this paper is structured as follows.
It starts with a short introduction to biometric verification in \autoref{sec:biometric-systems}.
Section~\ref{sec:template-protection} goes into the current state of the art of template protection.
It provides an overview of related work and defines the concept of a secure biometric template.
After the context of our work is given \autoref{sec:system-overview} provides a global overview of the secure biometric verification system we are introducing.
It covers the goals we are trying to achieve with the system, its architecture, and the security assumptions about its environment.
The paper continues with explaining the building blocks on which the system relies.
The classifier and the encryption scheme are covered in \autoref{sec:likelihood-ratio} and \autoref{sec:homomorphic-encryption}.
The exact functional description is covered in \autoref{sec:secure-verification}.
In \autoref{sec:security-argument} we prove the system's correctness and security.
The performance of the system is assessed in \autoref{sec:results}.
The paper concludes in \autoref{sec:conclusion} reflecting on our goals and looks ahead to future work.

\section{Biometric Systems}\label{sec:biometric-systems}
\begin{figure}[!h]
  \centering
   \includegraphics[width=.9\columnwidth]{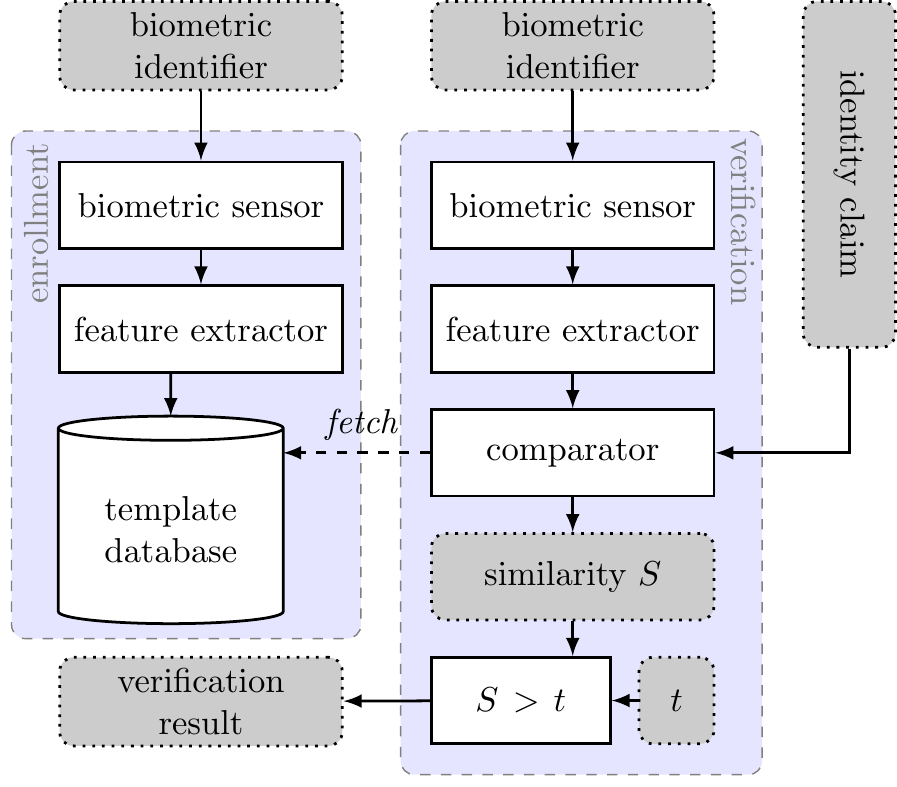}
  \caption{A schematic view of a generic biometric system. The left hand side models the enrollment process, while the right hand side models the verification process.
  The verification result shows whether the biometric identifier matches the identity claim, based on the similarity score $S$ and threshold $t$.}
  \label{fig:biometric-system}
\end{figure}

Biometric systems attempt to recognize a person based on his biometric characteristics~\cite{Jain2008}.
Therefore the system first learns about a person during an enrollment phase.
During the verification phase it checks whether a person matches a certain identity claim.
This section describes a generic biometric system as it is depicted in \autoref{fig:biometric-system}.

A system consists of multiple components which cooperate to produce a verification result; an identity claim gets accepted or rejected.
The first component is the biometric sensor which takes a biometric sample from a person.
The raw sensor data is processed by the feature extractor.
The \emph{features} are numeric representations of the identifier which together form a \emph{feature vector}.
During enrollment the system stores a feature vector as a \emph{template} in a database under a unique key for retrieval at a later time.

When a user makes an identity claim, the system captures his biometric identifier which is converted to a feature vector as a \emph{probe}.
The comparator checks if the probe and the template in the database form a matching identity.
As biometric identifiers may change over time and the sensor picks up noise, the probe will differ from the template and will never yield an exact match.
This makes the comparison a classification problem which is solved by the comparator by calculating a similarity score $S$.
This score expresses the confidence that two features belong to the same person.
If the score exceeds a certain threshold $t$ the identity claim gets accepted.

\begin{figure}[!h]
  \centering
   \includegraphics[width=.8\columnwidth]{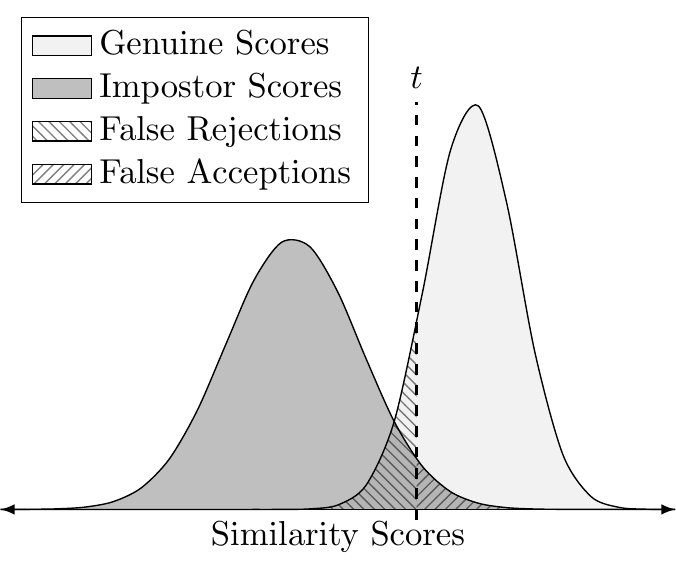}
  \caption{Classification Errors for a certain threshold $t$. Moving the threshold allows to accept higher \acp{FAR} for lower \acp{FRR}.}
  \label{fig:classification-error}
\end{figure}

Biometric matching is, as most classification problems, not perfect.
A genuine verification attempt may get rejected, or an impostor may mistakenly get accepted.
\autoref{fig:classification-error} shows the typical score distributions for genuine and impostor verifications.
The diagonally shaded areas denote the classification errors which depend on the threshold value.
The threshold can be tuned to shift the trade-off between a higher \ac{FAR} versus a lower \ac{FRR} and vice versa.

The accuracy of a biometric system is therefore expressed in the error rates given a certain threshold value~\cite{Jain2004}.
A good performing biometric system yields a low \ac{FAR} and a low \ac{FRR}.
One measure to express the performance of a biometric system is the \ac{EER} where the \ac{FAR} and \ac{FRR} are equal for a certain threshold.
An \ac{ROC} curve is used to express the systems accuracy in a graphical way.
This curve plots the \ac{FAR} against the \ac{GAR} for certain thresholds.
The \ac{GAR} is the percentage of correctly classified persons and equals $1 - $\ac{FRR}.

%

\section{Template Protection}\label{sec:template-protection}
Template protection schemes are designed to protect the biometric data when the biometric system gets compromised~\cite{Campisi2013}.
An attacker who gets a hold on a protected template should not be able to recover the original biometric data.
A \emph{secure template} should possess the following three properties as described by~\cite{Maltoni2009}.

\begin{enumerate}
  \item \emph{Non-invertibility}: It must be a computational hard task to recover the original biometric data from a secure template.
  The original template can be converted by applying a one-way function to it.
  One-wayness implies that the comparator also has no knowledge of the original biometric either and therefore should be able to operate on the transformed or encrypted data.
  \item \emph{Accuracy}: Transformation to a secure template may affect the systems matching precision negatively.
  If the transformation throws away too much information about the template the \ac{FAR} increases, which forms a vulnerability on its own.
  \item \emph{Diversity}: It must be possible to generate multiple versions of the same template.
  This property allows a template to be invalidated when it gets compromised and re-issue a template to grant the legitimate user access again.
  Another implication is that a user can use different `pseudonyms' (i.e.\ enrollments) for different systems which prevents traceability and protect his identity.
\end{enumerate}

Research in the area of template protection schemes can be split into two main categories following the research of~\cite{Nandakumar2008}: Feature Transformations and Biometric Crypto-Systems.
Apart from these two approaches a third category can be added which focusses on performing the template comparison in the encrypted domain.
Some examples of these techniques are described in the remainder of this section.

\subsection{Feature Transformations}
Secure templates based on Feature Transformations generate an alternative representation of the original biometric feature vector.
Comparing two feature vectors is done by applying the transformation function on both the template and the probe and perform the matching on the transformed data.
Known implementations of feature transformations are \emph{BioHashing}~\cite{Kong2006} and \emph{Cancelable Transformations}~\cite{Ratha2007}.

Transformations are based on a secret to create different representations.
A different secret yields a different representation which satisfies the diversity property of a secure template.

The challenge is to create a transformation function that is non-invertible and keeps the intra-user variations intact.
Studies show that feature transformations often create a trade-off between accuracy and non-invertibility~\cite{Lacharme2014,Nandakumar2008}.
This is unwanted as it makes it possible to create approximations of the biometric identifier from the secure template.

\subsection{Biometric Crypto-Systems}
This class of systems avoid the fuzzy matching process by finding the `stable' features in a biometric identifier.
These systems try to capture the same feature vector each time a user presents his identifier to the sensor.
By using error correcting techniques they try to reconstruct the features from the sensor data, and a piece of `helper data' which is stored during enrollment.
The helper data should not leak any data about the biometric identifier.
However, in practice this appears not to be the case and is it possible to extract information about the biometric identifier from it~\cite{Ignatenko2009}.

The stable feature vectors can be used as key in a cryptographic scheme, (e.g.\ AES).
Implementations of these systems can either be \emph{key-generating}~\cite{Dodis2008} or \emph{key-binding}~\cite{Nandakumar2007}.
In case of the former the identifier \emph{is} the key, while with the latter the key is chosen and encoded in the biometric using the helper data.

\subsection{Biometric Systems in the Encrypted Domain}
This class of systems provides true cryptographic security to protect the biometric data from leaking.
There are multiple options to perform biometric verification in the encrypted domain.
Homomorphic biometric systems are able to encrypt a biometric template while still being able to perform fuzzy matching on the encrypted data.
Another method is the use of \acp{GC} which allows two parties to evaluate a series of logical gates without learning its structure and only revealing the output.

Both approaches typically are very well in protecting privacy, but tend to be slow.
Computational complexity often forms a trade-off with accuracy.

One homomorphically based system shows a method for comparing IrisCodes which consist of 4096 binary features~\cite{Schoenmakers2007}.
Computing the similarity score of two feature vectors is done by calculating the hamming distance of the bit strings.
This operation can be done in the encrypted domain by using the Goldwasser-Micali scheme.
Initial results yield matching times of several minutes, which can be sped up running the matching algorithm on an approximation of the biometric templates.

Another, more generic homomorphic biometric system, is based on \acp{SVM}~\cite{Upmanyu2009}.
An \ac{SVM} is a powerful classification tool which projects feature vectors as points in a multi-dimensional space.
During enrollment it defines a hyperplane which divides the space into a part for genuine feature vectors and a part for impostors.
For classification it checks in which part the probe is placed.
This can be computed efficiently in the encrypted domain using RSA.
The authors report an \ac{EER} of 2\% at a matching speed of about a second per biometric comparison.

One method using \acp{GC} focusses on iris-code verification~\cite{Luo2012}.
It operates by implementing a logic circuit which computes the hamming distance over two bit strings representing the template and the probe.
The outcome is used in a circuit which performs the threshold comparison.
This method is fast (536ms) and yields an \ac{EER} of 1.4\%.

\subsection{Overview}
Following the definition of a secure template the biometric systems in the encrypted domain show the most promising results as shown in \autoref{tab:sectempcomp}.
The plus and minus signs in the table denote well or missing properties.
The ``\textpm'' denotes a property which is affected by a trade-off situation.

\begin{table}[!h]
\caption{Comparison of Secure Template Schemes}
\label{tab:sectempcomp}
\centering

\begin{tabular}{lccc}
\hline
  & Non-invertibility & Accuracy & Diversity \\
\hline
Feature Transform. & \textpm & - & + \\
Bio. Crypto-Systems & - & - & -\\
Sys. in the Encr. Dom. & + & \textpm & + \\
\hline
\textbf{Proposed System} & \textbf{+} & \textbf{+} & \textbf{+}
\end{tabular}

\end{table}

Feature transformations are either invertible or struggle to reach a high accuracy.
Some of these methods reveal privacy sensitive information about the biometric data, but solutions to overcome this have a negative effect on accuracy.

Biometric crypto-systems also have some open challenges.
The main problem is that these systems do not cope well with the diversity property.
Also, the error correction codes on which these schemes depend, need to be designed in such a way that they don’t leak information about the biometric data.
This appears to be hard according to some studies.

The approaches in the encrypted domain score well on both the non-invertibility as the diversity properties.
The main challenge is to balance computational complexity versus accuracy.

\section{System Overview}\label{sec:system-overview}
The previous section shows that the current state of the art of biometric systems in the encrypted domain is promising, but can be improved in terms of speed and accuracy.
This section provides an overview of a new biometric verification method es on privacy protection of its users, and aims to improve current solutions.

Our main goal is to create a biometric verification system which protects the users' privacy by not revealing any information about their biometric identifiers.
This goal should hold at any stage during the verification process, even if parts of the system are compromised.

The system requirements with regard to security are formulated as follows:
\begin{enumerate}
  \item The system does not reveal the biometric templates at any stage of the verification process.
  \item The biometric comparator does not learn the probes during the verification process.
  \item The system does not reveal the similarity score at any time during the verification process.
  \item The biometric comparator does not learn the verification result.
\end{enumerate}

To satisfy these requirements the system is constructed using homomorphic encryption techniques to perform the verification while securing the biometric data.

For a highly accurate system we rely on a likelihood ratio based classification system.
This is a generic classification method which can be applied on multiple biometric identifier types.
Studies show that this classifier behaves optimal in a sense that it yields the minimal \ac{FAR} for a given \ac{FRR} and vice versa~\cite{Bazen2004}.

First we describe the architecture and the components of the system.
Then we consider the environment the system operates in by defining the attacker model and listing the security assumptions we made.

\subsection{System Architecture}
\label{sec:architecture}

The system is constructed following the model of a generic biometric system in \autoref{sec:biometric-systems}.
We assume that it operates in a distributed environment where multiple clients are supported by a centralized server.
A communication channel exists between both parties to send and receive messages.

A client device embeds a sensor to capture a biometric sample from a user and takes the identity $u$ claim as input.
The device is able to pre-process the raw data and extracts the feature vector $\vec{p}$ from it.
The procedure $\mathsf{capture}()$ models this behavior and returns a tuple $(u, \vec{p})$.
The clients are called the \emph{sensor devices}.

The server is referred to as the \emph{verification service}.
The verification service is a database to secure biometric templates.
The procedure $\mathsf{FetchTemplate}(u)$ queries the database to find the template $T_u$ for identity $u$.

Feature comparison and matching is a joined effort of both parties which run a secure protocol to achieve that.
The result of the protocol is that only the client knows if the identity claim is accepted.

\subsection{Attacker Model}
\label{sec:attacker-model}

The attacker model defines what we are protecting against.
We assume that an adversary can be modeled by the semi-honest attacker model.
This assumption beholds that an attacker can gain full access to the system for observation, but will not alter its behavior.
Such an attacker is honest-but-curious and may try to gather as much information as possible, only based on his observations.

In case of a compromised party, the attacker gets full control about all the information of that part of system.

We assume that no collusion takes place.
This means that the sensor device and verification service do not exchange information to corrupt the system.
In other words it means that both systems may not get in control by the same attacker at the same time.

Finally, we also assume that enrollment takes place in an offline, secure setting.
In a real world situation this step should take place in a controlled environment where it is possible for the user to check that the device is not tampered with.

\section{Likelihood Ratio Based Classification}\label{sec:likelihood-ratio-based-classification}
The secure biometric verification system we describe in this paper is based on a Log-Likelihood Ratio Classifier.
This section describes how, and under which assumptions this classifier operates.
First we describe the standard likelihood ratio as it can be applied in a biometric setting.
Next we show a variant of the likelihood classifier by quantizing the features and the similarity scores.
These steps are required to apply an encryption layer over the system, protecting the biometric data.
Finally we describe the finite outcome space of the quantized classifier which we use for the system evaluation.

\subsection{Likelihood Ratio}
\label{sec:likelihood-ratio}

The likelihood function $\mathcal{L}(\omega | x)$ expresses the probability of observing a certain value $x$ given a certain class $\omega$.
The \ac{LR} $\Lambda(x)$ expresses how more likely it is that an observation $x$ belongs to a certain class.
The \ac{LR} is defined as:
\begin{equation}
  \Lambda(x) = \frac{\mathcal{L}(\omega | x)}{\mathcal{L}(\bar{\omega} | x)} = \frac{\Pr(x | \omega)}{\Pr(x | \bar{\omega})}
\end{equation}

In the biometric setting the input consists of a probe and a template feature $p$ and $t$.
Instead of denoting wether an input is from a certain class (i.e.\ user), it calculates the \ac{LR} of two features being from the same user $\mathcal{L}(\text{same} | p, t)$, or being from different users $\mathcal{L}(\overline{\text{same}} | p, t)$.
The \ac{LR} in the biometric setting becomes:
\begin{equation}
  \Lambda(p, t) = \frac{\mathcal{L}(\text{same} | p, t)}{\mathcal{L}(\overline{\text{same}} | p, t)} = \frac{f_g(p, t)}{f_b(p, t)}
\end{equation}

In this equation $f_g$ and $f_b$ are respectively the \acp{PDF} of the genuine and the background distribution.

We assume that biometric features can be modeled as samples from a Gaussian distribution $T$.
A feature can then be modeled by a user specific component $M$ and a noise component $N$ such that $M + N = T \sim \mathcal{N}(0, 1)$.

The class mean $\mu_u \in M \sim \mathcal{N}(0,\sigma^2_b)$ is the distinctive value to classify a feature to a user $u$.
The parameter $\sigma^2_b$ is referred to as the \emph{between} user variance.

Noise can be modeled as Gaussian distribution $N \sim \mathcal{N}(0, \sigma^2_w)$.
The parameter $\sigma^2_w$ is called the \emph{within} user variance (i.e.\ intra-user variance).
Noise is modeled per feature, independent from the user~\cite{Chen2008}.
This means that noise has the same effect on a feature for each user.
Though, different features have different noise behavior, described by the value of $\sigma^2_w$.
This models `good' features which are less influenced by noise and are easier to classify than `bad' features.

The \ac{PDF} $f_g$ is a bivariate Normal distribution $\mathcal{N}(\left(\begin{smallmatrix}0\\0\end{smallmatrix}\right), \mat{\Sigma})$.
The covariance matrix $\mat{\Sigma}$ is defined by:
\begin{equation}
  \mat{\Sigma} = \begin{pmatrix}
  \E(T_pT_p) & \E(T_pT_t) \\
  \E(T_tT_p) & \E(T_tT_t) \end{pmatrix} = \begin{pmatrix}
  1 & \sigma^2_b \\
  \sigma^2_b & 1 \end{pmatrix}
\end{equation}
In this equation $T_p$ and $T_t$ are the distributions of the inputs $p$ and $t$.
As the inputs are features from the same person they share the same user specific component but have different noise components (resp. $M + N_p$ and $M + N_t$).

The \ac{PDF} $f_b$ takes two independent variables as input as $p$ and $t$ are inputs from different users.
Therefore it is possible to write the joint probability as the multiplication of the two single probabilities $f_t(x)f_t(y)$.
Here is $f_t$ the \ac{PDF} of the total feature distribution $T$.

With the probability density functions known, the likelihood ratio can be computed.
For computational convenience the \ac{LLR} $\lambda$ can be used:

\begin{equation}
  \lambda(p, t) = \tfrac{1}{2}\left((p^2+t^2) - \left( \begin{smallmatrix}p & t\end{smallmatrix}\right) \mat{\Sigma}^{-1}\left( \begin{smallmatrix}p \\ t\end{smallmatrix}\right)\right) - \tfrac{1}{2}\ln(|\mat{\Sigma}|)
\end{equation}

%

To use the \ac{LLR} it is assumed that it is possible to extract uncorrelated features from a biometric identifier.
This allows to compare two feature vectors by summing the individual feature comparisons.
A comparator $\mathcal{C}$ to calculate the similarity score of $k$-dimensional feature vectors is defined as:
\begin{equation}
  \mathcal{C}(\vec{p}, \vec{t}) = \sum_{i=0}^{k-1} \lambda(p_i, t_i)
\end{equation}

\subsection{Feature Quantization}
\label{sec:feature-quantization}

In order to lower the complexity of the \ac{LLR} comparator we propose a way to precompute the outcomes.
To limit the infinite possible outcomes in the continuous case we quantize the features over $2^b$ equiprobable bins $B$.
Equiprobability means that an arbitrary feature observation is just as likely to land in any of those bins.

We define the quantization function $q: T \mapsto B$ to determine in which bin a feature is placed.
The functions $u(\beta)$ and $l(\beta)$ respectively determine the upper and lower bounds of a bin $\beta \in B$.
Then $u'(x)$ and $l'(x)$ are respectively defined as $(u \circ q)(x)$ and $(l \circ q)(x)$.

In the quantized case the \ac{LR} can be redefined as $\Lambda'$.
The numerator becomes the probability of an observation in a certain area, given that the features originate from the same person.
The denominator becomes the probability of observing an arbitrary combination of two features in that area:
\begin{multline}
  \Lambda'(p,t) =\\
  \frac{\Pr(l'(p) < T_p < u'(p),  l'(t) < T_t < u'(t)  |\text{same})}{\Pr(l'(p) < T_p < u'(p),  l'(t) < T_t < u'(t))}
\end{multline}

The quantized version of the \ac{LLR} $\lambda'$ can then be defined by integrating over the \acp{PDF}.
\begin{equation}
  \begin{aligned}
  \lambda'(p,t) = &\ln(\int_{l'(p)}^{u'(p)}\int_{l'(t)}^{u'(t)} f_g(p, t) \dif t\dif p) -\\
  &\ln(\int_{l'(p)}^{u'(p)}\int_{l'(t)}^{u'(t)} f_b(p, t)\dif t\dif p)
  \end{aligned}
\end{equation}

This definition of the \ac{LLR} can be used to precompute the similarity scores of all possible observations of a quantized feature.
The results can be organized in a in a $2^b \times 2^b$ lookup table $\mathcal{T}_{b, \rho}$.

\begin{equation}
    \mathcal{T}_{b, \rho} = \begin{pmatrix}
      s_{0,0} & s_{0,1} & \cdots & s_{0,2^b-1} \\
      s_{1,0} & s_{1,1} & \cdots & s_{1,2^b-1} \\
      \vdots  & \vdots  & \ddots & \vdots  \\
      s_{2^b-1,0} & s_{2^b-1,1} & \cdots & s_{2^b-1,2^b-1} \end{pmatrix}
\end{equation}
In this table are $x, y \in B$ and $s_{x,y} = \lambda^*(x,y)$; the \ac{LLR} with integral limits defined by $u$ and $l$.
The subscript $\rho$ denotes the characteristic between user variance $\sigma^2_b$

We use $\mathcal{T}_{b, \rho}^{(x)}$ to denote the $x^{\text{th}}$ row of a lookup table.

\subsection{Score Quantization}
\label{sec:score-quantization}

The lookup table $\mathcal{T}_b$ consists of similarity scores $s_{x,y} \in \mathbb{R}$.
A second quantization step is required to convert the scores to integer values in order to protect these number using an encryption layer.
Therefore we define $q_{\Delta} : \mathbb{R} \mapsto \mathbb{N}$ where $q_{\Delta}(s)$ is a uniform quantization function with step size $\Delta$.

\subsection{Score Distribution}
\label{sec:score-domain}
In the continuous case the number of possible comparison scores is infinite.
Due to the quantization steps the outcomes of the comparison are limited.
The quantized score distribution  of a single quantized feature is the set of the values  in a lookup table.
\begin{equation}
  S = \{s_{x,y} | x,y \in B\}
\end{equation}
The score domain $[\min(S), \max(S)]$ describes the boundaries of possible scores.

We can also consider the score distribution of a biometric identifier described by a $k$-dimensional feature vector.
The total score distribution $\mathbb{S}$ is then defined by the convolution of the single feature distributions $S_0 \ast S_1 \ast \dots \ast S_{k-1}$.
The score domain of $\mathbb{S}$ is defined by $[\sum^{k-1}_{i=0} \min(S_i), \sum^{k-1}_{i=0} \max(S_i)]$.

\section{Homomorphic Encryption}\label{sec:homomorphic-encryption}
To protect the templates we use ElGamal encryption as it possesses all the cryptographic properties we need~\cite{ElGamal1985}.
This section gives a short review on the cryptographic scheme and discusses the properties used in the proposed system.

The security of ElGamal is provided by the \ac{DDH} assumption about the underlying cyclic group.
The \ac{DDH} implies that it is computationally hard to solve a logarithm in a cyclic group.
This is known as the \ac{DLP}.

Let $G$ be a cyclic group of size $p$ generated by $g$.
The private key $a$ is a random number between 1 and the order of the group.
The public key $h$ is constructed by raising the generator to the power $a$:
\begin{equation}
  \langle g \rangle = G, \quad |G| = p, \quad a \in_R (0,|G|), \quad h = g^a
\end{equation}

The encryption function $E$ to encrypt message $m \in G$ yields a tuple $(c_1, c_2)$ is defined by:
\begin{equation}
  E(m) = (g^r, \quad mh^r) \quad \text{with} \quad r \in_R (0,|G|)
\end{equation}
The random $r$ makes ElGamal probabilistic such that different encryptions of the same message yield different ciphertexts.
This property also makes ElGamal \ac{INDCPA} secure under the \ac{DDH} assumption.

For notation we denote an encrypted value with double brackets $\ct{{\cdot}}$.
In case $m$ is a vector or matrix the elements of $\ct{m}$ are encrypted component-wise.

Any party in possession of the secret key $a$ can evaluate the decryption function $D$ which is defined as:
\begin{equation}
  \begin{aligned}
  D(c_1, c_2) = c_1^{-a}c_2 &= (g^r)^{-a}(m(g^a)^r) \\
                            &= m(g^{r-a}g^{a+r}) = m
  \end{aligned}
\end{equation}

\subsubsection{Additive Homomorphism}
\label{sec:additive-homomorphism}

Homomorphic cryptographic schemes allow us to make computations in the encrypted domain.
ElGamal is multiplicative homomorphic as $\ct{m}\ct{m'} = \ct{mm'}$ which is illustrated below:
\begin{equation}
  \begin{aligned}
  (c_1, c_2)(c'_1, c'_2) &= ((g^r)(g^{r'}), &(mh^r)(m'h^{r'})) \\
  &= (g^{r+r'}, &mm'h^{r+r'})
  \end{aligned}
\end{equation}
During secure biometric verification we want to sum encrypted similarity scores from the feature comparators in the encrypted domain.
It is possible to use ElGamal in an additive homomorphic mode by encoding the messages as exponent of the generator $g^m$~\cite{Cramer1997}. This yields:
\begin{equation}
  \ct{g^{m_1}} \ct{g^{m_2}} = \ct{g^{m_1 + m_2}}
\end{equation}

The downside of this approach is that after decrypting the ciphertext the message can not be retrieved without solving the \ac{DLP} $g^m$.
However, it is trivial to spot wether $m$ equals zero as $m = 0 \iff g^m = 1$.
We will make use of this fact to achieve an efficient secure comparison protocol in \autoref{sec:secure-biometric-verification}.

\subsubsection{Secret Sharing}

No single party in the system should be able to decrypt a ciphertext on his own as it may leak the templates.
Therefore it is possible to set up a threshold variant of ElGamal~\cite{Desmedt1989}.
As there are only two parties, which both should comply to a decryption we can simply split the secret key additively in two random shares: $a = a_1 + a_2$.

A party, which possesses one of the key shares $a_1$ is able to perform a partial decryption:
\begin{equation}
  D_1(c_1, c_2) = (c_1^{-a_1}, \quad c_2) = (c_1', c_2')
\end{equation}
The function $D_1$ yields a partially decrypted ciphertext which we denote by single brackets $\pdct{{\cdot}}$.
A partially decrypted ciphertext does not leak any information as it is indistinguishable from an ElGamal encryption of a random message under the public key $g^{a_2}$.

A second decryption step $D_2$, using the second part of the key retrieves the message.
\begin{equation}
  \begin{aligned}
  D_2(c_1', c_2') &= (c_1'^{-a_2})(c_2') = ((g^r)^{-a_1})^{-a_2}(m(g^a)^r) \\
  &= (g^{-r(a_1+a_2)})(m(g^{ar})) = m
  \end{aligned}
\end{equation}

\section{Secure LLR Based Biometric Verification}\label{sec:secure-verification}
This section combines the building blocks from the previous sections in order to create a system to perform a secure biometric verification using the quantized log-likelihood based classifier.
The main objective is that the server gets no knowledge about the biometric data.

To achieve this we run a two party protocol such that table lookups can be done in an oblivious fashion such that both parties can not tell which values are selected.
Protecting the lookup values is required as it directly points to an $x$ and $y$ value which encodes the probe and template input.

We also want to protect the similarity score to prevent a hill climbing attack.
Observing the score allows an attacker to modify its input feature vector such that it gets closer to the original template with each iteration.

Finally we want to make sure that the verification result is only learned by the client and not by the server.

\subsection{Secure Biometric Enrollment}
\label{sec:secure-biometric-enrollment}

The first step, prior to verification, is to enroll new users to the system.
Enrollment takes place on a sensor device which captures a biometric identifier and creates a secure template from it.
A secure template is created from raw, unprotected biometric data.
Therefore we should assume that the enrollment of new users happens in an offline, secure and fully controlled environment.

The enrollment process starts with constructing a secure template.
Therefore we first start with a definition of a template in this system.
Note that, considering a lookup of a similarity score $s_{x,y}$ in a quantized comparator, the $x$ and $y$ value are respectively the enrollment observation and the verification observation.
A template can thus be seen as the first argument of a lookup in a table.
It selects the row where the similarity score can be found during verification.

For a biometric identifier described by a $k$-dimensional feature vector $k$ lookup tables $\mathcal{T}_{b, \rho_0}, \mathcal{T}_{b, \rho_1}, \dots, \mathcal{T}_{b, \rho_{k-1}}$ can be constructed.
A quantized feature vector $\vec{p}$ defines the rows to select during enrollment.
The template $\mat{T}_u$ can be constructed by augmenting the corresponding rows from the lookup tables.
\begin{equation}
  \mat{T}_u = \left((\mathcal{T}_{b, \rho_0}^{(p_0)})^\top | (\mathcal{T}_{b, \rho_1}^{(p_1)})^\top | \dots | (\mathcal{T}_{b, \rho_{k-1}}^{(p_{k-1})})^\top \right)^\top
\end{equation}

After the sensor device constructed the template from the enrollment data it encrypts it using the public key of the system.
The encrypted template $\ct{\mat{T}_u}$ is send over to the verification service which stores the template under key $u$ for later retrieval.

The lookup tables should be considered public knowledge to anyone who is able to retrieve the variance vector $\langle \rho_0, \rho_1, \dots \rho_{k-1} \rangle$ which describes the feature distributions.
The sole purpose of encryption is to hide which rows are selected from these tables.

\subsection{Secure Biometric Verification}
\label{sec:secure-biometric-verification}

\begin{figure}[!h]
  \centering
  \includegraphics[width=\columnwidth]{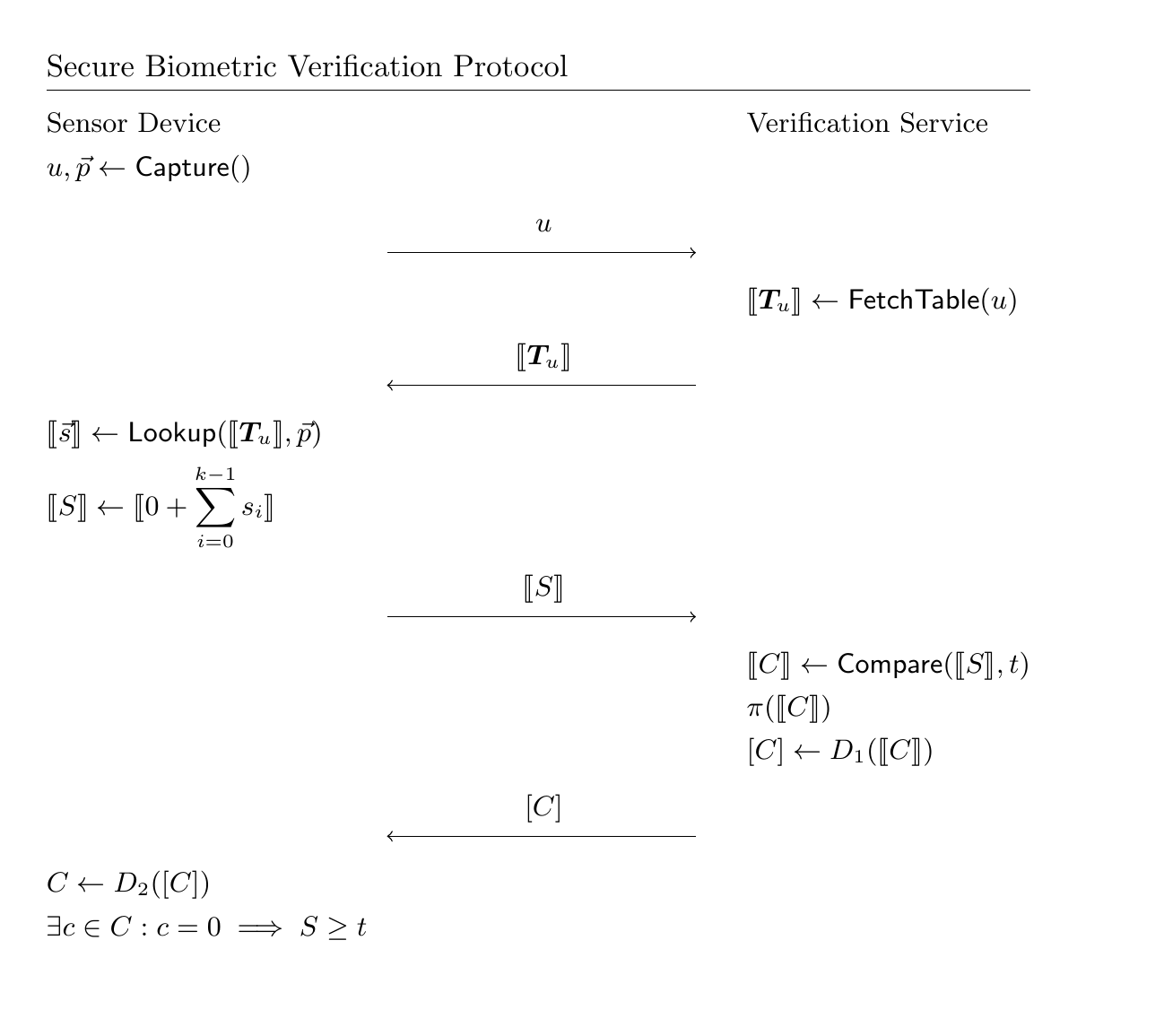}
  \caption{Secure Verification Protocol}
  \label{fig:verification-protocol}
\end{figure}

The verification protocol consists of a comparison and matching stage.
The first stage compares a biometric probe to a secure template and yields an encrypted similarity score $\ct{S}$.
The second stage determines if there is a match by securely comparing the score to a threshold $t$.
Each stage completes in a single communication round between the sensor and the verification service.
A schematic view of the protocol is given in \autoref{fig:verification-protocol}.

\subsubsection{Secure Comparison}

The comparison protocol starts with a call to $\mathsf{capture}()$ at the sensor which yields a quantized probe vector $\vec{p}$ and an identity claim $u$.
The identity claim is sent over the the verification service which uses $u$ to fetch the corresponding secure template $\ct{\mat{T}_u}$ from its database.
The template is sent to the sensor.

The sensor uses $\vec{p}$ and $\ct{\mat{T}_u}$ to perform a $\mathsf{lookup}$ for each feature.
The rows in $\ct{\mat{T}_u}$ are partial lookups which have been created during the enrollment stage.
Each feature $p_i$ defines the column in the $i^{\text{th}}$ row of $\ct{\mat{T}_u}$ which contains the similarity score $\ct{s_i}$.
This score equals the score $s_{x,y} \in \mathcal{T}_{b, \rho_i}$ and expresses the likelihood of enrollment feature $x$ and probe feature $y$ being from the same person.

To come to a cumulative (i.e.\ vector) similarity score $\ct{S}$ the sensor sums the feature scores $\ct{0 + \sum_{i=0}^{k-1} s_i}$.
The addition of zero randomizes the outcome of the sum.
This prevents an attacker who gets hold of the encrypted template to guess the elements which yield the same value as $\ct{S}$ as this would leak the probe.

\subsubsection{Secure Matching}

The second stage tries to determine wether the similarity score exceeds a certain the threshold $\ct{S > t}$.
There is no direct homomorphic equivalent for the comparison operator, but it is possible to check for equality by checking wether the decryption of $\ct{S-t}$ equals $0$.

The score is an element of the total score distribution $S \in \mathbb{S}$.
The distribution is finite and discrete due to the quantization of the comparator as shown in \autoref{sec:score-domain}.
Therefore it is possible to check if the score $S$ has a value $t \le S \le \max(\mathbb{S})$ by calculating $\ct{S-t-i}$ for all $0 \le i \le \max(\mathbb{S})-t$.

This stage of the protocol starts with the sensor sending $\ct{S}$ to the verification service.
The verification service then computes the encrypted result set $\ct{C}$ which equals $\ct{\{r(S-t-i) | \forall 0 \le i \le \max(S)-t, r \in_R [1,|G|] \}}$.
The procedure $\mathsf{compare}()$ performs this computation based on an encrypted score and a threshold value.

To protect the values in the result set we multiplicative blind the elements with a random value $r$.
Finally we remove any ordering in the set by applying a random permutation $\pi(\ct{C})$.

To check for a match the verification service creates a partial decryption $\pdct{C}$ and sends it back to the sensor.
The sensor decrypts the result set $C$ and concludes that $S$ is larger then the threshold value if, and only if the result set contains a $0$-value.

\section{Correctness and Security Analysis}\label{sec:security-argument}
This section proves that the proposed system is actually secure and does not leak any information in the semi-honest model.
This is done by systematically checking the correctness and security for both the enrollment and the verification protocol.

Correctness shows that the proposed protocols work as intended under the assumption that both parties act according to the description.
We show that each procedure in the system outputs correct results.

To prove security of the protocols we follow the ``real-vs.-ideal'' framework~\cite{Goldreich2009}.
This framework shows that a protocol can be modeled in the ideal situation by performing all computations at a trusted third party given the input parameters.
Then the framework states that all malicious intents in the real protocol can be simulated in the ideal model.
The Composition Theorem~\cite[Theorem 7.3.3]{Goldreich2009} allows to evaluate the security of each step in a protocol individually to make statements about the protocol as a whole.

\subsection{Correctness}

\subsubsection{Enrollment Protocol}
\begin{itemize}

  \item We assume that the $\mathsf{capture}$ procedure at the sensor device is implemented correctly and yields the intended results: the $k$-dimensional feature vector $\vec{p}$ and the identity claim $u$.

  \item Template construction is described in \autoref{sec:secure-biometric-enrollment}.
The template $\mat{T}_u$ is a $k$ by $2^b$ matrix where each row equals the ${p_i}^{\text{th}}$ row of lookup table $T_{b, \rho_i}$.
Each element in the template gets encrypted using the ElGamal cryptographic scheme to get $\ct{\mat{T}_u}$.
This does not alter the values of the template by definition of the encryption function shown in \autoref{sec:homomorphic-encryption}.

  \item The database stores the exact value of $\ct{\mat{T}_u}$ under the key $u$ for retrieval at verification time.
\end{itemize}

\subsubsection{Verification Protocol}
\begin{itemize}
  \item We make the same assumption as in the enrollment protocol that the $\mathsf{capture}$ procedure is implemented correctly and yields a $k$-dimensional quantized feature vector $\vec{p}$ and identity claim $u$.

  \item $\mathsf{fetchTemplate}$ queries the database for the template under key $u$.
The result is the exact output of the template which is stored during enrollment.
If that procedure is performed correct, the output $\ct{\mat{T}_u}$ consists of an encrypted $k$ by $2^b$-matrix where the $i^{\text{th}}$ row equals the ${p_i}^{\text{th}}$ row from $\mathcal{T}_{b, \rho_i}$.

  \item Given the values in $\vec{p}$ $\mathsf{lookup}$ selects element $p_i$ from the $i^{\text{th}}$ row of $\ct{\mat{T}}$ which equals an encryption of the feature similarity score $s_{p_i, i} \in \mathcal{T}_{b, \rho_i}$.
The elements $\ct{s_{p_i, i}}$ are summed using the additive homomorphic property of ElGamal to compute $\ct{S}$.

  \item The verification service receives the encrypted similarity score $\ct{S}$ which it uses to compute the matching set $\ct{C}$ using the $\mathsf{compare}$ procedure.
The $\mathsf{compare}$ algorithm tests for equality by calculating $\ct{C} = \{r\ct{S-t-i} | 0 \le i \le \max(\mathbb{S}) - t \}$ with random $r \in [1, |G|]$.
The homomorphic subtraction and the scalar multiplication are correct by the definition of ElGamal in \autoref{sec:homomorphic-encryption}.
If $S \ge t$, then $r\ct{S-t-i} = \ct{0}$ for a certain $i$.
All other values are encryptions of random numbers.
The procedure returns a partial decrypted, permuted version of the matching result: $\pdct{C}$.
Permuting the elements of $\pdct{C}$ only changes the order but not its values.

  \item The final step decrypts the values of $\pdct{C}$ which results in a shuffled set of the values $C = \{r(S-t-i) | 0 \le i \le \max(S)-t \}$.
\end{itemize}

\subsection{Security}

\subsubsection{Enrollment Protocol}

\paragraph{Compromised Sensor}
We assume that during enrollment the sensor is placed in a controlled offline environment where no tampering with the device occurs during template construction.

\paragraph{Compromised Verification Service}
The server receives an encrypted template $\ct{\mat{T}_u}$.
The template consists of encrypted rows from different lookup tables.
Due to the \ac{INDCPA} security property of ElGamal the server is not able to distinguish an encryption of a certain row from random data.
Also, due to the equiprobable distributed features, each row is as likely to be included in the template as any other row which makes it impossible to make an educated guess.

\subsubsection{Verification Protocol}

\paragraph{Compromised Sensor}
First we consider the situation where the sensor is compromised.
During the first round of the protocol the sensor receives $\ct{T_u}$, a matrix of ElGamal encrypted messages.
Based on the \ac{INDCPA} security of ElGamal we can claim that an adversary at the sensor can not distinguish between the elements in the matrix, and therefore does not learn anything about it.
As the used cipher is threshold based, the adversary at the sensor only has a part of the decryption key, which makes it impossible to decrypt the matrix on its own.

After the second round in the protocol the sensor receives $\pdct{C}$, a vector of partial decrypted messages.
Using secret key of the sensor it can recover $C$.
The values in $C$ are blinded by an unknown randomization factor $r$ in the $\mathsf{compare}()$ algorithm.
As $0\cdot r = 0$ the only value which is revealed to the sensor is $0$ which is the intended outcome.
Any information introduced by the sequential construction of $C$ in $\mathsf{compare}()$ is removed by returning a random permutation of the result.

\paragraph{Compromised Verification Service}
The second part of the security argument involves the evaluation of the situation where the verification service is compromised.
During the first round of the protocol the verification service receives an unencrypted identity claim $u$, which is by design of the verification process.
The claim does not contain any information on the biometric, so the requirements are still met.

During the comparison round of the protocol the verification service gets $\ct{S}$.
This number is encrypted under Threshold ElGamal, and therefore $\ac{INDCPA}$ secure.
No information about $S$ is revealed.

\subsection{Secure Biometric Template}

In this section we stress that the biometric templates as defined in this system satisfy the non-invertibility and diversity properties of a secure template.

A secure template consists of an encrypted matrix where each row is selected from a lookup table $\mathcal{T}_{b, \rho}$.
First we note that each row is as likely to be selected as any other row due to the equiprobable feature quantization.
This makes it impossible to make an educated guess which row is encrypted.
Second, we note that is impossible to distinguish between rows due to the $\indcpa$ property of ElGamal.
This property makes that identical scores in the lookup table have different representations in the encrypted form.
This rules out the option to identify a row from the lookup table based on the encrypted representation.

These two characteristics make that no information is leaked from the secure template.
Given that it is infeasible to break the ElGamal cipher without knowledge of the decryption key we claim that the templates in this system satisfy the non-invertibility property.

The diversity property is trivial by the possibility to re-randomize ciphertexts in ElGamal.
This allows us to generate complete new representations without changing the content of the template.

\section{Experimental Results}\label{sec:results}
In order to make any claims about the performance of our systems we created a proof of concept implementation.
This section focusses on performance in terms of accuracy and speed.

Both the feature and the score quantization steps influence the accuracy of the system.
To show that these effects are minimal we conduct a small experiment which benchmarks the quantized comparator to the continuous (i.e. optimal) case.
As the encryption layer does not influence the accuracy of the system we can evaluate the quantization effects on its own.

The quantization parameters and the feature selection have an effect on the processing speed of the system.
To get a good intuition about how the parameters influence the system we simulate various operational settings to evaluate the system performance.

\subsection{Accuracy Assessment}
\label{sec:accuracy-assessment}

In order to show that the effects introduced by quantizing the comparator are minimal we created a set of \ac{ROC} curves to asses the precision loss.
Quantization depends on two parameters $b$ and $\Delta$.

The feature quantization parameter $b$ yield a denser lookup table for higher values.
A denser lookup table gives a better approximation of the continuous \ac{LLR}.
The downside of a denser lookup table is that it requires more space to store, which also impacts the processing time.

The score quantization parameter $\Delta$ influences the scoring scale.
Lowering $\Delta$ creates a more accurate scale, but drastically increased the score domain which negatively impacts the performance.

\begin{figure}[!h]
  \centering
  \includegraphics[width=\columnwidth]{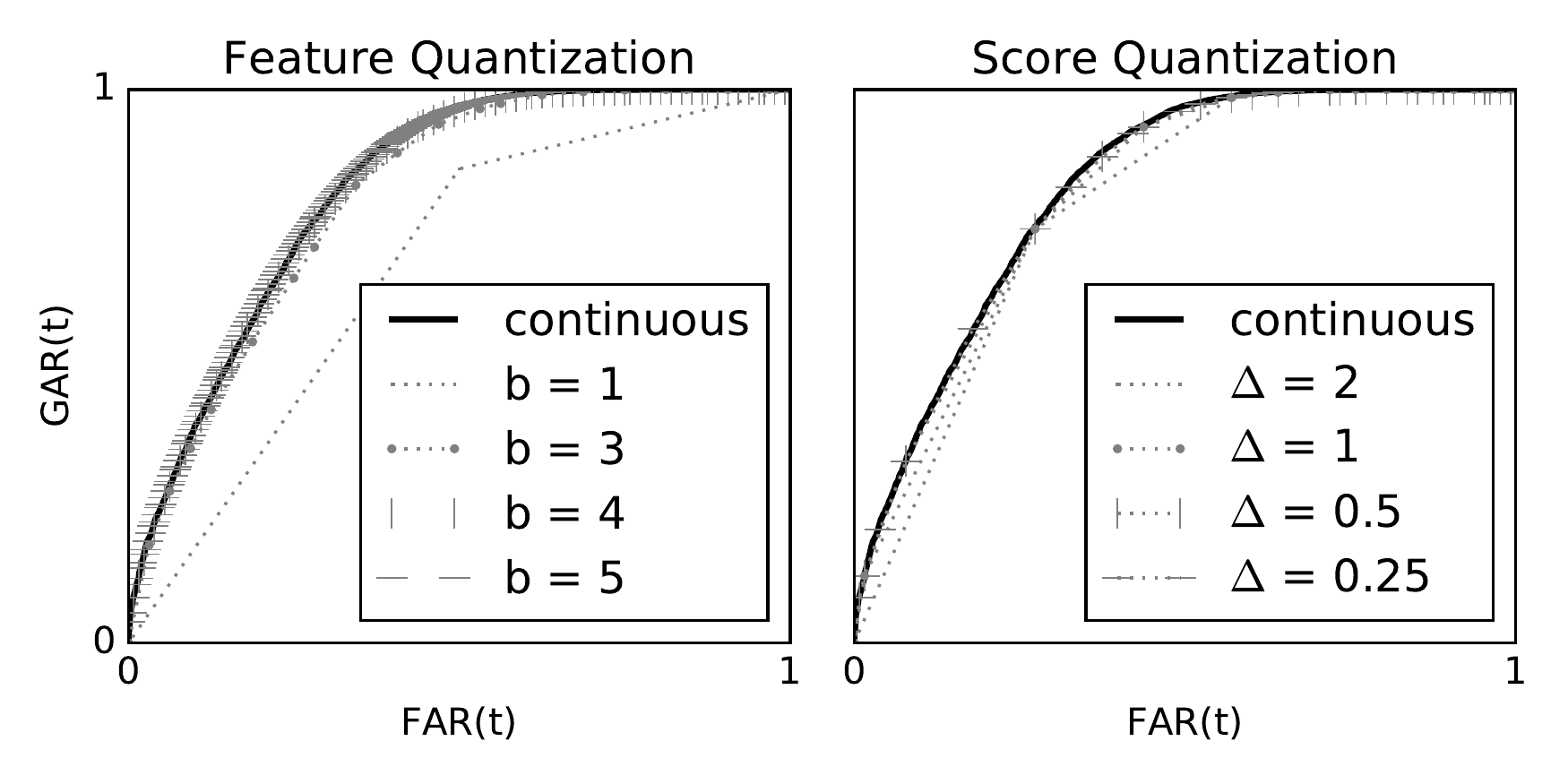}
  \caption{\ac{ROC} curves illustrating the accuracy impact of quantization. Comparing the continuous case to a feature with $\sigma^2_b = 0.9$.}
  \label{fig:quantization-roc}
\end{figure}

\autoref{fig:quantization-roc} shows the \ac{ROC} curves for various quantization parameters.
The left figure benchmarks different feature quantization parameters against the optimal comparator.
It shows that the quantized \ac{ROC} curves approach the continuous case really quick to a point where it makes no sense to make the lookup table any denser.
This break-even point is reached at $b=4$ for a feature with characteristic $\sigma^2_b = 0.9$.
This point is reached faster for lower between class variances.
Raising the variance requires higher values of $b$.
However, in reality features of such precision do not exist.

The right plot in \autoref{fig:quantization-roc} shows how the score quantization influences the precision of the comparator considering the same feature with a feature quantization of $b=4$.
It shows that the score quantization barely impacts the accuracy and that a $\Delta = 1$ is a reasonable value to use.

Based on these results the quantized \ac{LLR} comparator can be used as a good approximation of the continuous comparator.

\subsection{System Speed Assessment}

\subsubsection{Implementation and Runtime Environment}

To run some basic experiments we implemented the system and some measurement tools in Python $3.5.1$.
The ElGamal cryptographic layer is implemented on a group structure defined on elliptic curves.
Elliptic curves provide a speed up compared to integer group structures and require smaller group orders for the same level of security.
For the implementation of elliptic curve cryptography we rely the SECCURE cryptographic library.
The selected curve is named `secp112r1' which is a curve over a 112-bit prime field.
This curve is mainly chosen because of its small group size and fast performance on arithmetic operations.

The experiments are performed on a MacBook Pro 2.8GHz Intel Core i7, 16GB memory, Intel Iris 1536MB graphical chip, with a 512GB SSD.
This system is able to perform an ElGamal elliptic curve encryption in 2.96 ms, measured over 100 samples.
Decryptions require 938 $\mu$s on average over 1000 samples.
Homomorphic additions take 12 $\mu$s (10000 samples) and scalar multiplications are done in 1.95 ms (1000 samples).

\subsubsection{Parameter Selection}

The bottleneck of the system lies in the score comparison protocol.
This is by far the slowest step of the protocol as it requires the computation of $\alpha = \max(\mathbb{S}) - t$ encrypted numbers.
A large score domain yields a larger value for $\alpha$.

The following parameters influence the score domain.
\begin{enumerate}
  \item the number of features $k$.
  \item the quality of the feature $\rho$.
  \item the feature quantization parameter $b$.
  \item the score quantization parameter $\Delta$.
  \item the threshold $t$.
\end{enumerate}
In an operational setting the feature parameters $k$ and $\rho$ are implied by the biometric identifier.
However, in this experimental setting the features are simulated.
For this experiment we select 3 feature sets with different properties which could represent real world situations (See \autoref{tab:feature-sets}).

\begin{table}[!h]
\caption{Feature Sets}
\label{tab:feature-sets}
\centering

\begin{tabular}{l | c l}
\hline
id & $k$ & Features $\vec{\rho}$  \\
\hline
fs1 & 21 & $\langle 0.7, 0.71, \dots, 0.89, 0.9 \rangle$ \\
fs2 & 20 & $\langle 0.8, 0.8, \dots, 0.8, 0.8 \rangle$ \\
fs3 & 12 & $\langle 0.7, 0.7, 0.7, 0.7, 0.8, 0.8, 0.8, 0.8, 0.9, 0.9, 0.9, 0.9 \rangle$ \\
\hline
\end{tabular}
\end{table}

The experiment is conducted with 6 quantization settings which we apply on each feature set.
The quantization parameters are chosen based on the observations of the accuracy experiment.
The number of comparisons depends on the threshold $t$.
To come to a fair comparison $t$ is set to the value for which the comparator reaches the \ac{EER}.
We express its speed in terms of the number of comparisons $\alpha$.
A single number comparison requires one homomorphic addition $\ct{S-t'}$, one scalar multiplication $\ct{r(S-t')}$, a decryption step at the verification service $\pdct{r(S-t')}$ and a decryption at the sensor device.
\autoref{tab:timings} shows the results of measuring the processing time for various values of $\alpha$.
The performance scales linear with the size of $\alpha$ allowing us to do around 250 comparisons per second.

\begin{table}[!h]
\centering
\caption{Processing Time for $\alpha$ Threshold Comparisons}
\label{tab:timings}
\begin{tabular}{l || c | c | c | c | c | c | c}
$\alpha$ & 10 & 20 & 30 & 40 & 50 & 60 & 80 \\
\hline
time (in ms) & 40 & 80 & 130 & 170 & 210 & 250 & 330
\end{tabular}
\end{table}

\subsubsection{Performance Results}

\begin{figure*}[h!]
  \centering
  \includegraphics[width=0.9\textwidth]{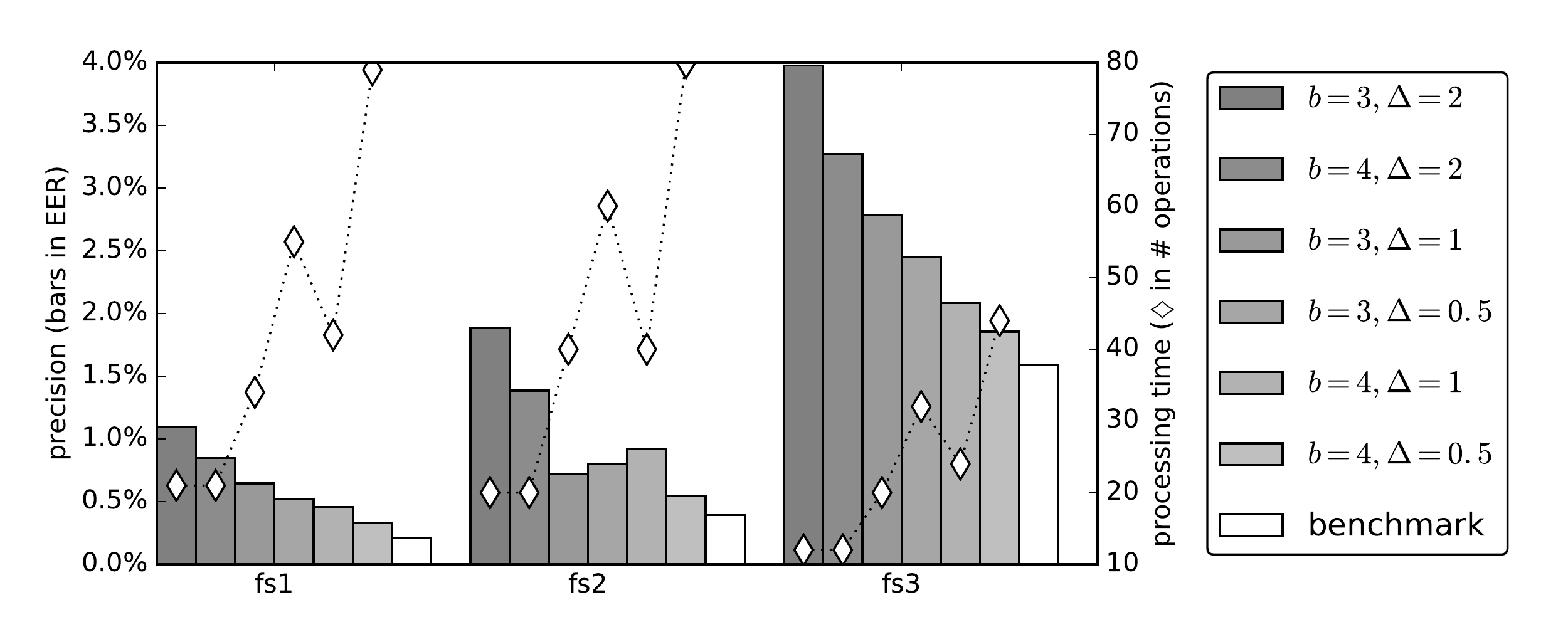}
  \caption{System performance with several quantization parameters shown on 3 feature sets (fs1, fs2 and fs3). The bars show the precision of the system expressed in EER on the left axis. The diamonds show the processing time expressed in numbers of threshold comparisons on the right axis. The benchmark shows which accuracy can be achieved with the \ac{LLR} comparator when no quantization takes place.}
  \label{fig:performance-results}
\end{figure*}

\autoref{fig:performance-results} shows the results of the 18 experiments.
The grouped bars represent the results for various feature selections.
The different shades of grey represent the quantization parameter sets.
The height of the bars express the accuracy of the system as \ac{EER}.
For each feature set we add the \ac{EER} in continuous setting as benchmark which we consider as the optimal case for the \ac{LLR} comparator.
The diamonds show how well the system performed in terms of speed, expressed in the number of threshold comparisons $\alpha$.

The graphs clearly show the trade-off between accuracy and processing time as better accuracy values are associated with larger values for $\alpha$.
It also shows that larger values of $k$ (fs1 and fs2) yield better results then poorer feature selections (fs3).

We note that the the feature quantization parameter $b$ barely influences the outcome of the experiment in number of threshold comparisons.
Yet, raising this value improves the accuracy of the system which is in tune with the observations in the accuracy experiment.

Observing the score quantization parameter $\Delta$, we note a bigger influence on the system.
Lowering $\Delta$ exponentially increases the required number of threshold comparisons, which slows the system down.
Though, if time is less of an issue, it is possible to gain some accuracy here.

Overall the timings are very low and well within the range of practical use cases.
This claim holds, even in the worst-case parameter selections.
On top of that it shows the \ac{EER} does not deviate a lot from the optimal case where no quantization is applied.

\section{Conclusion and Future Work}\label{sec:conclusion}

We presented a novel secure biometric verification system based on a \ac{LLR} comparator and ElGamal homomorphic cryptography.
In order to apply the encryption layer to the system we quantized the biometric data and adapted the \ac{LLR} comparator accordingly.

Measurements show that the goals in terms of accuracy and speed are well met.
Highly accurate secure biometric verifications (with an \ac{EER} of 0.3\%) can be done in about 330 milliseconds.
Accepting slightly less accurate settings lowers this time to 150 milliseconds or even faster (\textless 50 milliseconds) if accuracy is less important.

With regard to privacy and security we conclude that the templates satisfy all three properties of a secure template as defined in \autoref{sec:template-protection}.
No information can be extracted from the biometric templates due to the encryption layer.
Yet, the templates can be used for verification without decrypting the biometric data.
Measurements show that this can be done in a highly accurate way compared to the optimal case.
Diversity is achieved by using different keys for different applications.

\subsection{Future Work}
Our research evaluation is based on simulated biometric data.
This shows very promising results but may not be a good representative for real world behavior as it is hard to estimate good feature parameters.
A follow up research should test the system on real biometric datasets.

In terms of security we are aware of the relative small number of possible inputs for the system.
This is inherent to biometric data and not a shortcoming of this implementation.
An attacker with a lot of time and computational power should be able to brute force this space to gain access to the system.
In practice this is not a problem as we can limit the number of false requests to the verification service.
This effectively limits the number of adversarial trials.
Another way to mitigate this is to raise the number of features which describe the identifier.
Though, it might be hard as an identifier might not contain enough entropy as the features should be independent.
Combining multiple biometric identifiers can probably solve this issue.
We see this as interesting future work.


Finally, we designed this system with biometric verification in mind.
Though, this approach might be applicable in different fields where classification of encrypted data is required.
Further research might focus on generalizing the system for other applications.
As part of this generalization it should be investigated if data distributions, other then Gaussian distributed biometric data, are equally suitable for this classification method.

\bibliographystyle{IEEEtranS}
\bibliography{refs}

\end{document}